\documentclass[twocolumn,preprintnumbers]{revtex4}
\usepackage{amsmath,amssymb,graphics,epsfig,subfigure}
\usepackage{color}
\usepackage[colorlinks,linkcolor=red,anchorcolor=red,citecolor=green]{hyperref}
\usepackage{setspace}
\begin{document}

\thispagestyle{empty}

\begin{center}

\title{The fine micro-thermal structures for the Reissner-Nordstr\"{o}m black hole}

\author{Zhen-Ming Xu$^{}$\footnote{E-mail: xuzhenm@nwu.edu.cn}, Bin Wu$^{}$\footnote{E-mail: binwu@nwu.edu.cn (Corresponding author)},
        and Wen-Li Yang$^{}$\footnote{E-mail: wlyang@nwu.edu.cn}}

\affiliation{ $^{2}$Institute of Modern Physics, Northwest University, Xi'an 710127, China\\
$^{1}$School of Physics, Northwest University, Xi'an 710127, China\\
$^{3}$Shaanxi Key Laboratory for Theoretical Physics Frontiers, Xi'an 710127, China\\
$^{4}$Peng Huanwu Center for Fundamental Theory, Xian 710127, China}

\begin{abstract}
Based on the idea of the black hole molecule proposed in [Phys. Rev. Lett. 115 (2015) 111302], in this paper, by choosing the appropriate extensive variables, we have solved the puzzle whether the molecules of the Reissner-Nordstr\"{o}m black hole is an interaction or not through the Ruppeiner thermodynamic geometry. Our results show that the Reissner-Nordstr\"{o}m black hole is indeed an interaction system that may be dominated by repulsive interaction. More importantly, with the help of a new quantity, thermal-charge density, we describe the fine micro-thermal structures of the Reissner-Nordstr\"{o}m black hole in detail. It presents three different phases, the {\em free}, {\em interactive} and {\em balanced} phases. The thermal-charge density plays a role similar to the order parameter, and the back hole undergoes a new phase transition between the {\em free} phase and {\em interactive} phase. The competition between the {\em free} phase and {\em interactive} phase exists, which leads to the extreme behavior of the temperature of the Reissner-Nordstr\"{o}m black hole. For extreme Reissner-Nordstr\"{o}m black hole, the whole system is completely in the {\em interactive} phase. What is more significant is that we provide the thermodynamic micro-mechanism for the formation of the naked singularity of the Reissner-Nordstr\"{o}m black hole.
\end{abstract}

\maketitle
\end{center}

\section{Motivation}
With the advent of the first black hole's image and the observation of more and more gravitational wave events, black hole physics is playing an increasingly prominent role in our understanding of gravity. It is now generally believed that the black hole hides clues about how to unify general relativity and quantum mechanics. The emergence of the thermodynamics of black holes provides a powerful way to explore the secret of black holes\cite{Hawking1975,Bardeen1973,Bekenstein1973,Hawking1983}. Especially the introduction of extended phase space\cite{Kastor2009,Dolan2011} enriches the research contents of black hole thermodynamics, such as the analogy between the charged AdS black hole and van der Waals fluid\cite{Kubiznak2012}, the application of the Maxwell equal area law\cite{Kubiznak2012,Spallucci2013}, the microscopic analysis of black hole phase transition\cite{Wei2015,Zangeneh2016,Wei2019a,Wei2019b} and etc. A large number of research results have revealed that black holes have microstructures. Nevertheless, describing the microscopic behavior of black holes remains a huge challenge in the current gravitational problem.

Recently, a new concept, i.e., black hole molecule, has been proposed, which provides a new perspective for studying the micro-mechanism of black holes phenomenologically\cite{Wei2015,Wei2019a,Wei2019b,Wei2019c}. This new scheme is mainly based on the Ruppeiner thermodynamic geometry which introduced a Riemannian metric structure to represent the thermodynamic fluctuation theory\cite{Ruppeiner1995}. In order to make the thermodynamic theory has a mathematical description method in geometric sense, Weinhold\cite{Weinhold1975} introduced a thermodynamic metric $g^{W}_{\mu\nu}=\partial^2 U/(\partial X^{\mu}\partial X^{\nu})$ in terms of the internal energy to achieve this goal. However, the Weinhold geometry seems physically meaningless in the context of purely equilibrium thermodynamics. Subsequently, Ruppeiner further developed the geometric theory of thermodynamics, and put forward a theory later called Ruppeiner thermodynamic geometry. It is based on the second-order partial derivative structure of the entropy with respect to other thermodynamic quantities, and has the physically meanings in the fluctuation theory of equilibrium thermodynamics. The components of the inverse Ruppeiner metric gives second moments of fluctuations. The above two geometries are proved to be conformal equivalent with the conformal factor $1/T$. Afterwards, the Ruppeiner geometry is widely used to explore some micro-information of the black hole thermodynamic system and ordinary fluid thermodynamic system. While the Weinhold geometry has no such effect.

Thermodynamic curvature is the most important physical quantity in the Ruppeiner thermodynamic geometry. For some better understood statistical mechanical models in  ordinary thermodynamics, the sign of thermodynamic curvature can qualitatively reflect some information about the character of the molecular interaction for a thermodynamic system. There is an empirical observation that a positive (or negative) thermodynamic scalar curvature implies a repulsive (or attractive) interaction, and a vanishing thermodynamic scalar curvature corresponds to no interaction\cite{Ruppeiner2014,Ruppeiner2008}. Meanwhile its absolute value can reflect the strength of molecular interaction in some sense. The big absolute value of the thermodynamic curvature implies strong interaction and the small one corresponds to weak interaction\cite{Miao2018}. For black hole systems, because there does not exist a complete theory of quantum gravity, the exploration on the microscopic structure of black holes is bound to some speculative assumptions. Due to the well-established black hole thermodynamics, as an analogy analysis and a primary description of the micro-behavior of black holes, one can think that the empirical observation mentioned above can provide some information about interactions of black holes phenomenologically or qualitatively \cite{Cai1999,Liu2010,Xu2020,Niu2012,Wang1910,Guo2019,Wang2018,Li2016,Ghosh2020,Zangeneh2018}.

For the Reissner-Nordstr\"{o}m (RN) black hole, we have already known its thermodynamic behaviors and the temperature of the black hole shows a maximum. However, a series of studies\cite{Aman2003,Aman2006a,Aman2006b} have suggested that the RN black hole is a non-interacting system by means of the Ruppeiner thermodynamic geometry. On the other hand, the literature\cite{Mirza2007} pointed out that the results of the RN black hole should be reduced from those of the Kerr-Newmann-AdS black holes, and the coordinate space of the thermodynamic geometry of the RN black hole adopted in the papers\cite{Aman2003,Aman2006a,Aman2006b} maybe is incomplete. Their results indicated that the RN black hole is an interacting system. So there are divergences on whether there is interaction between molecules of the RN black hole. Intuitively, as an independent thermodynamic system, the coordinate space of the thermodynamic geometry of the RN black hole should be complete and its metric of the thermodynamic geometry is well (we will see it in the analysis later). We speculate that the reason of non-interacting between RN black hole molecules in the papers\cite{Aman2003,Aman2006a,Aman2006b} deserves further investigation.
Moreover, as a charged black hole, there should be interaction between its molecules, like the electromagnetic interaction. Hence we firmly believe that the RN black hole is a complete and interacting system. These are the main motivations of this study.

Now in present paper, by choosing the appropriate extensive variables, we show that the black hole has repulsive interaction. More importantly, with the help of a new quantity, called thermal-charge density, we describe the fine micro-thermal structures of this black hole in detail. It has three different phases, the {\em free} phase, {\em balanced} phase, and {\em interactive} phase. More meaningful is that we provide the thermodynamic micro-mechanism for the formation of the naked singularity of the RN black hole. Throughout this paper, we adopt the units $\hbar=c=k_{_{B}}=G=1$.

\section{The fine micro-thermal structure}\label{sec2}
For the RN black hole, its mass $M$, charge $q$ and the horizon radius $r_{\pm}$ satisfy the following relationships
\begin{equation}\label{horizon}
r_{\pm}=M \pm \sqrt{M^2-q^2}.
\end{equation}

One can see that the black hole has two horizons, and $r_{+}>r_{-}$. Hence the event horizon is marked as $r_h=r_{+}$. When $M=q$, two horizons merge into one. This is the extreme black hole, and the condition $M\geq q$ is also known as the Bogomol'nyi boundary.

The basic thermodynamic properties of the RN black hole take the following forms in terms of the event horizon radius $r_h$\cite{Aman2003,Aman2006a,Aman2006b},
\begin{eqnarray}
\text{Internal energy}&:&U=M=\frac{r_h}{2}+\frac{q^2}{2r_h},\label{energy}\\
\text{Temperature}&:&T=\frac{1}{4\pi r_h}-\frac{q^2}{4\pi r_h^3},\label{temperature}\\
\text{Entropy}&:&S=\pi r_h^2.\label{entropy}
\end{eqnarray}

In previous discussions, one can think that the internal energy $U$ of the black hole is a function of entropy $S$ and charge $q$, i.e., $U=U(S, q)$. Correspondingly, the first law of thermodynamics is written as $d U=T d S+\varphi dq$, where the $\varphi=q/r_h$ is the electrostatic potential. In the coordinate space $\{S, q\}$, the RN black hole is a non-interacting system\cite{Aman2003,Aman2006a,Aman2006b}. Furthermore, the study\cite{Mirza2007} suggested that the RN black hole is an interacting system according to the degeneration of results of the Kerr-Newmann-AdS black holes and indicated that the coordinate space $\{S, q\}$ is not complete. Our current view is that the RN black hole is a complete and interacting system. It is likely that the core of the divergence among previous works on interaction between molecules of the RN black hole is the selection of the coordinate space for the Ruppeiner thermodynamic geometry.

Inspired by the result of the study \cite{Dehyadegari2017} in which authors used $q^2$ as an independent thermodynamic quantity to analyze the critical behavior and  microscopic structure of the charged AdS black hole for the first time and the black hole also exhibits the van der Waals-type phase transition behavior in the new phase space, without extended phase space. Similar phenomena also appear in the Gauss-Bonnet gravity as well as in higher dimensional spacetime\cite{Yazdikarimi2019}. In this paper, we find that the coordinate space $\{S, Q\}$
is appropriate to eliminate the above divergence, where the new thermodynamic quantity thermal-charge $Q$ is taken as $q^2$.
Then we have
\begin{equation}\label{thermalcharge}
Q=q^2, \qquad \Psi=\left(\frac{\partial U}{\partial Q}\right)_S=\frac{1}{2r_h},
\end{equation}
and the first law of thermodynamics and Smarr relation can be written in terms of the thermodynamic quantities mentioned above as follows:
\begin{equation}\label{fsrealtion}
d U=T d S+\Psi dQ, \qquad U=2(TS+\Psi Q).
\end{equation}

Now we make some comments on the above description:
      \begin{itemize}
        \item For the phrase ``the coordinate space'' mentioned above, let us make a simple explanation. In classical mechanics, a system with $n$ degrees of freedom has a $2n$ dimensional phase space. In analogy, in a thermodynamic system, the phase space should contain both the conjugated extensive and intensive quantities (or generalized coordinates and their conjugate generalized forces). Here the phase space of the RN black hole in our scheme should be $\{T, \Psi, S, Q\}$. For the theory of thermodynamics geometry, we do it in a space of generalized coordinates, like as $\{T, \Psi\}$, $\{S, Q\}$, $\{T, Q\}$ and $\{S, \Psi\}$ for the RN black hole.
        \item The RN black hole has a zero thermodynamic curvature in the coordinate space $\{S,q\}$ indicating that it is a non-interacting system. This seems to be inconsistent with our intuitive understanding of the RN black hole. In order to cure this problem, we want to introduce a new scheme to achieve this goal. By observing the form of solution of the RN black hole and the expressions of its related basic thermodynamics quantities, we find that the parameter of charge appears in the form of $q^2$. Therefore, for simplicity, we regard $Q=q^2$ as an independent thermodynamics quantity. Its conjugate quantity is $\Psi=1/(2r_h)$ with the simplest form (it is only a function of the horizon radius $r_h$, while the electric potential $\varphi=q/r_h$ is a function of $q$ and $r_h$.). With the help of such a pair of new quantities, as shown in later, we can obtain that the RN black hole is indeed an interacting system in the new coordinate space $\{S,Q\}$.
        \item The new physical quantity $\Psi$ conjugated with $Q$ does not depend on the charge $q$, which is different from the electric potential $\varphi$. But their contributions to the internal energy of the system are the same, i.e., $2Q\Psi=q\varphi$. On the other hand, according to the idea of the black hole micromolecule introduced by Wei and Liu in Ref.[9], we can see clearly that $\Psi$ is equal to the number density of black hole micromolecules in the natural units, which is the inverse of the specific volume defined in Ref.[7].
      \end{itemize}

In the new coordinate space $\{S, Q\}$, we can show that the RN black hole is indeed an interaction system and can also give some fine microstructures of the RN black hole completely from the thermodynamic point of view, with the help of the Ruppeiner thermodynamic geometry. The metric can be written in the internal energy form
\begin{equation}\label{rmetric}
g_{\mu\nu}=\frac{1}{T}\frac{\partial^2 U}{\partial X^{\mu}\partial X^{\nu}},
\end{equation}
where $X^{\mu}$ represents some independent thermodynamic quantities. For the RN black hole, $X^{\mu}$ are $S$ and $Q$. The line element takes the form
\begin{equation}\label{line}
d l^2=\frac{1}{C_{_Q}}d S^2+\frac{2}{T}\left(\frac{\partial T}{\partial Q}\right)_S dS dQ+\frac{1}{T}\left(\frac{\partial \Psi}{\partial Q}\right)_S d Q^2,
\end{equation}
where $C_{_Q} :=T(\partial S/\partial T)_{_Q}=2S(S-\pi Q)/(3\pi Q-S)$ and we have used the Maxwell relation $(\partial T/\partial Q)_{_S}=(\partial \Psi/\partial S)_{_Q}$ based on the first law of thermodynamics of the RN black hole (see Eq.~(\ref{fsrealtion})). The line element $d l^2$ measures the distance between two neighbouring fluctuation states in the state space.

Naturally we can obtain the thermodynamic scalar curvature
\begin{equation}
R=\frac{1}{S-\pi Q}. \label{curvature}
\end{equation}
According to the above formula, we can directly see that the curvature is not zero, which means that the RN black hole is indeed an interacting system. This is also consistent with our understanding of the electromagnetic interaction between charged black hole molecules. Meanwhile it also verifies the rationality of the coordinate space $\{S,Q\}$ we choose. In the light of Eqs.~(\ref{temperature}) and~(\ref{entropy}), because of the non-negative requirement of temperature for the black hole, we have $R>0$ which may be related to the information of repulsion interaction between black hole molecules for the RN black hole, i.e. electromagnetic repulsion interaction.

Now we make some further explanations of the above results which we obtain.

A series of studies \cite{Aman2003,Aman2006a,Aman2006b} suggested that the RN black hole has a zero thermodynamic curvature. While in the literatures \cite{Mirza2007,Quevedo2008,Shen2007}, they reported that the RN black hole has a non-zero thermodynamic curvature. About the puzzle on whether the thermodynamic curvature of the RN black hole is non-zero, our present results are qualitatively consistent with that in the literatures \cite{Mirza2007,Quevedo2008,Shen2007}. The results show that the RN black hole has a non-zero and positive thermodynamic curvature.

In Ref.\cite{Mirza2007}, authors calculate the thermodynamic curvature of the Kerr-Newmann-AdS black hole in the coordinate space $\{S,q,J\}$, and then takes the zero-limit of cosmological constant $\Lambda$ and angular momentum $J$ to obtain the non-zero thermodynamic curvature of the RN black hole. It is a possible way to cure the problem that the RN black hole is a non-interacting system. Meanwhile they point out that the coordinate space of the thermodynamic geometry of the RN black hole itself is incomplete. However, we think that the evidence to the above conclusion, that the coordinate space of the thermodynamic geometry of the RN black hole is incomplete according to the degenerated result of the Kerr-Newmann-AdS black hole, seems insufficient and questionable. In our current paper, instead we treat the RN black hole as an independent thermodynamic system. We propose an other way that a new thermodynamic coordinate $Q=q^2$ is adopted to achieve a non-zero thermodynamic curvature for the RN black hole, without the help of the degenerated case of the Kerr-Newmann-AdS black hole.

In Ref.\cite{Quevedo2008}, the geometrothermodynamics scheme has been proposed to obtain a non-zero thermodynamic curvature of the RN black hole. In Ref.\cite{Shen2007}, authors used the redefinition of the internal energy of the system to realize a non-zero thermodynamics curvature for the RN black hole (there is a sign difference in the Christoffel symbol between this paper and our work in the calculation of thermodynamic curvature).

Next we introduce a new quantity called thermal-charge density
\begin{equation}\label{thdensity}
\sigma=\frac{3\pi Q}{S}.
\end{equation}

The main reasons why we introduce the thermal-charge density are as follows:
      \begin{itemize}
        \item Because the entropy of a black hole is proportional to its horizon area, we call this quantity a kind of density (specifically, the surface density).
        \item With the help of the thermal-charge density $\sigma$, we can clearly describe the microstructure of the RN black hole. Meanwhile the new physical quantity plays a role similar to the order parameter.
      \end{itemize}

Obviously, $0<\sigma\leq 3$ and at $\sigma=3$ the temperature equals zero (extreme black hole). The term $3\pi Q$ on the numerator implies the existence of electromagnetic repulsion interaction. While the denominator $S$ denotes the disorder of molecules in a black hole system. The larger the entropy $S$, the stronger the disorder of the system. That is to say, the more violent the irregular thermal motion of molecules is. In the next analysis, we consider the case of fixed thermal-charge $Q$. Then the temperature Eq.~(\ref{temperature}) and the thermodynamic scalar curvature Eq.~(\ref{curvature}) for the RN black hole can also be expressed as rescaled temperature $T_r$ and rescaled thermodynamic scalar curvature $R_r$
\begin{equation}\label{reducetr}
\begin{split}
T_r &=4\pi\sqrt{Q}T=\sqrt{\frac{\sigma}{3}}\left(1-\frac{\sigma}{3}\right),\\
R_r &=\pi Q R=\frac{\sigma}{3-\sigma}.
\end{split}
\end{equation}

Hence we obtain the three different situations:
\begin{itemize}
  \item When $\sigma=1$, the temperature $T_r$ reaches its maximum $T_r=T_{max}=2/(3\sqrt{3})$ and the thermodynamic scalar curvature $R_{\sigma=1}=1/2$.
  \item When $0<\sigma<1$, we have $T_r: 0\rightarrow T_{max}$ and $R_r: 0 \rightarrow 1/2$.
  \item When $1<\sigma\leq 3$, we have $T_r: T_{max}\rightarrow 0$ and $R_r: 1/2 \rightarrow +\infty$. Especially at $\sigma=3$, we have $T_r=0$ and $R_r=+\infty$. This situation corresponds to the extreme black hole.
\end{itemize}

\begin{figure}
 \begin{center}
 \includegraphics[width=70mm]{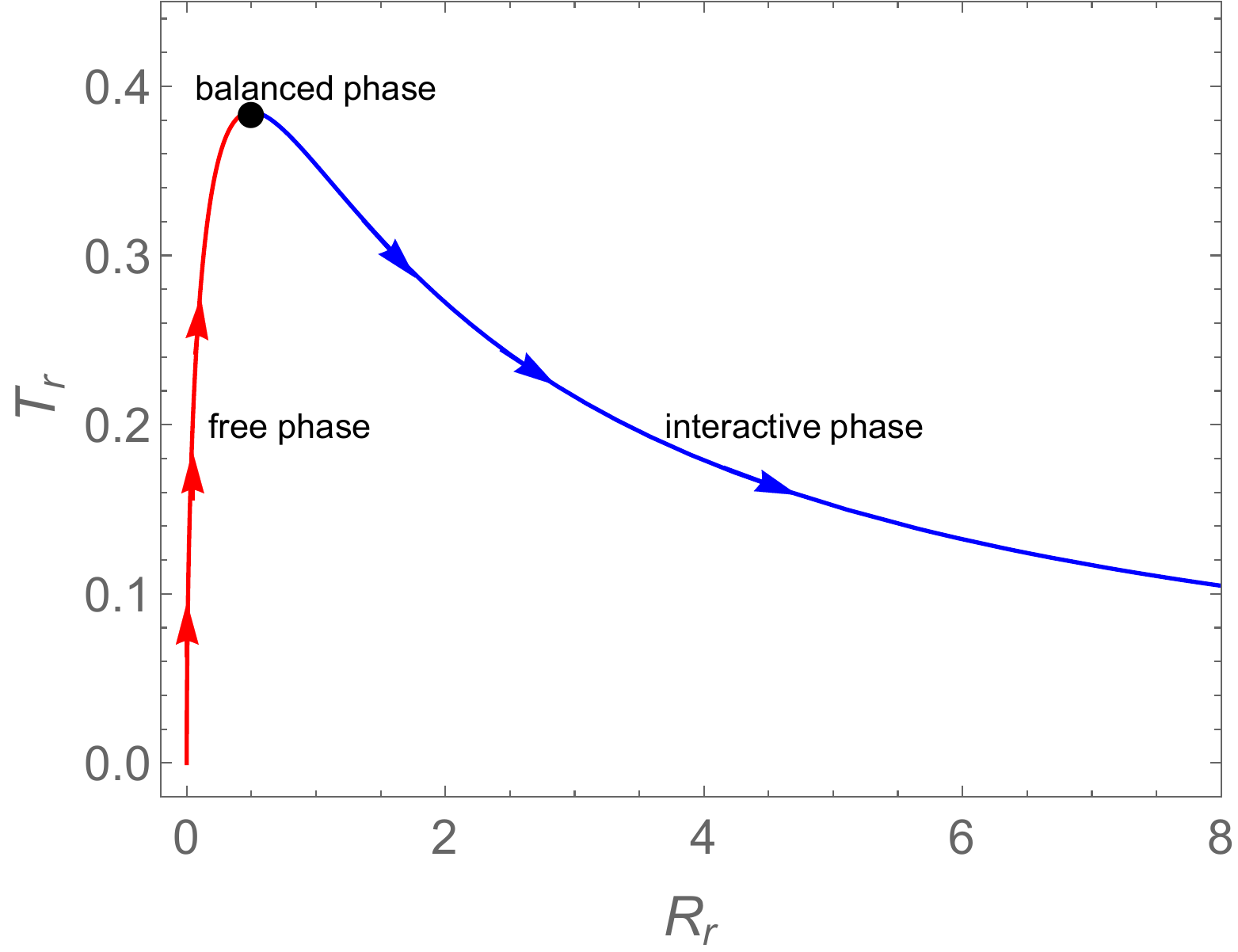}
 \end{center}
 \caption{The diagram of the rescaled temperature $T_r$ with respect to the rescaled thermodynamic scalar curvature $R_r$ and arrows indicate the direction in which the thermal-charge density increases.}
 \label{fig1}
 \end{figure}

Here, along the direction of increasing thermal-charge density $\sigma$, we show the curve of temperature $T_r$ and curvature $R_r$ in FIG. \ref{fig1}.
We can clearly see that there exists an inflection point in the $T_r-R_r$ plot, i.e., the maximum point. At the same time, we can observe the existence of three different phases.

{\bf Case 1:} {\em free} phase-- With the increasing of thermal-charge density $\sigma$ from $0$ to $1$, according to Eq.~(\ref{thdensity}), we can clearly see that the irregular thermal motion of molecules plays a major role. We call this situation the {\em free} phase. On the other hand, the thermodynamic curvature increases with the increasing of thermal-charge density, which maybe indicate that the repulsive interaction between black hole molecules is increasing. The repulsive interaction between a large number of molecules leads to the existence of {\em interactive} phase which will suppress the {\em free} phase. So there will be a competitive relationship between the two phases. But at the interval $\sigma \in (0,1)$, the {\em free} phase dominates the whole black hole system, and then the temperature of black hole is increasing.

{\bf Case 2:} {\em balanced} phase-- At $\sigma=1$, the competition between the {\em free} phase and {\em interactive} phase is balanced, which makes the temperature of black hole reach maximum. This is also the microscopic mechanism of the existence of the maximum temperature of the RN black hole.

{\bf Case 3:} {\em interactive} phase-- With the increasing of thermal-charge density $\sigma$ from $1$ to $3$, the thermodynamic curvature rapidly increases, which maybe imply that the repulsive interaction between black hole molecules is increasing rapidly. At this time, the {\em interactive} phase dominates the whole black hole system, which leads to a decrease in the temperature of black hole.

When $\sigma=3$, we have $T_r=0$ and $R_r=+\infty$. This is the extreme RN black hole, i.e. $M=q$. This moment a strong repulsive interaction dominates between black hole molecules. The whole black hole system is completely in the {\em interactive} phase and the temperature equals to zero. Now we consider the near-extreme circumstances, i.e. $\sigma\rightarrow 3$, we have the following relation with the help of Eq.~(\ref{reducetr}),
\begin{equation}
R_r=\frac{1}{T_r}-\frac32-\frac{3T_r}{8}-\frac{T_r^2}{2}+\mathcal{O}(T_r^3).
\end{equation}
Hence we obtain
\begin{equation}\label{rt}
\lim_{T_r\rightarrow 0} T_r R_r=1,\quad \text{or} \quad \lim_{M\rightarrow q} T R=\frac{1}{4\pi^2 q^3}.
\end{equation}
That is to say, when the Bogomol'nyi boundary is saturated, although the temperature $T$ tends to zero and the thermodynamic scalar curvature $R$ tends to infinite, the combination $T R$ of the two is a finite value.

When the thermal-charge density exceeds $3$, this situation corresponds to the case $M<q$ for the RN black hole. In the light of Eq.~(\ref{horizon}), it is clear that the event horizon of black hole is not exist, and the RN black hole becomes a naked singularity. From our current point of view, at that moment of $\sigma>3$, according to Eq.~(\ref{curvature}) or Eq.~(\ref{reducetr}), we can clearly see that the thermodynamic scalar curvature tends to be negative infinite, which implies that a strong attraction interaction dominates between black hole molecules. This causes the whole system to collapse into one point, that is, the naked singularity. In this way, we provide the thermodynamic micro-mechanism for the formation of the naked singularity.

At the end of this section, let's see whether the black hole has a phase transition. We know that the micro-mechanism of phase transition is the result of the competition between the interaction among the molecules that make up the system and the irregular thermal motion of the molecules themselves. Using the Ruppeiner thermodynamics geometry, we can reflect the micro-mechanism of the black hole phase transition to a certain extent. According to the theory of the molecular thermal motion and the significance of the thermodynamics curvature \cite{Wei2019a,Miao2018}, we approximately think that the temperature can be regarded as a measure of the irregular thermal motion, while the thermodynamic curvature can be regarded as a measure of the molecular interaction.

According to FIG. 2, we can clearly see that there is a gap between the {\em free} phase and the {\em interactive} phase, where the ``gap-1'' indicates that the black hole system is dominated by the thermal motion, while the``gap-2'' means that the system is dominated by the interaction. In other words, the RN back hole undergoes a new phase transition between the {\em free} phase and {\em interactive} phase and the phase transition point is exactly at the {\em balanced} phase, i.e., the thermal-charge density $\sigma=1$.
\begin{figure}
 \begin{center}
 \includegraphics[width=70mm]{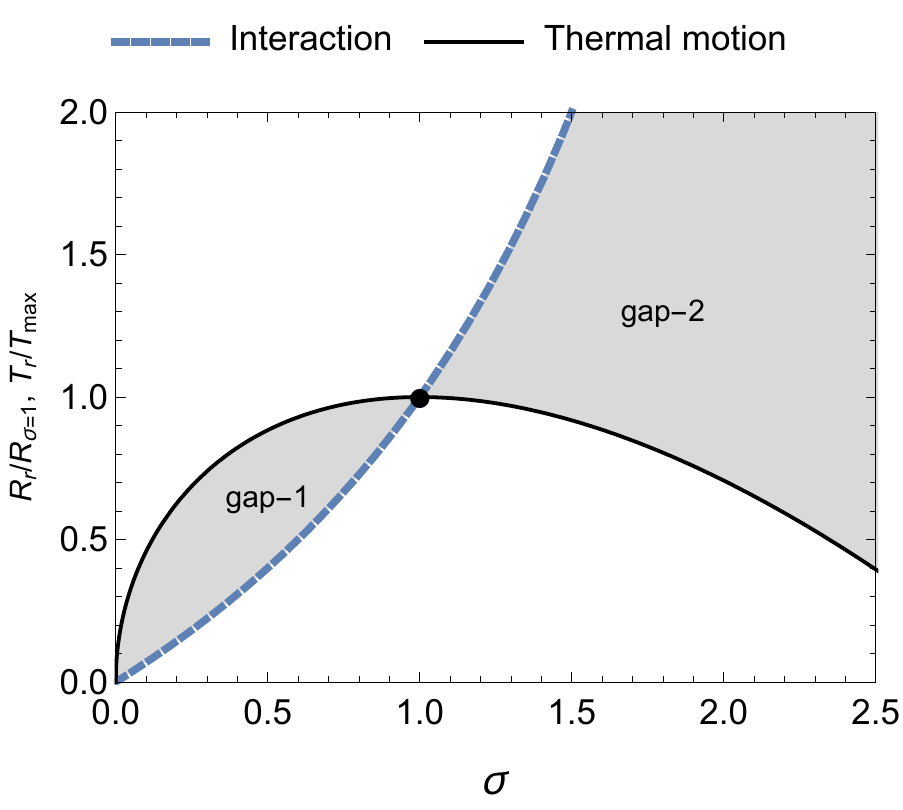}
 \end{center}
 \caption{The behaviors of the dimensionless measurement $T_r/T_{\text{max}}$ of the thermal motion and the dimensionless measurement $R_r/R_{\sigma=1}$ of the interaction with respect to the thermal-charge density $\sigma$ respectively.}
 \label{fig2}
 \end{figure}

\section{Summary and Discussion}
By choosing the appropriate extensive variables, we show that the RN black hole maybe have repulsive interaction. More importantly, with the help of a new quantity, i.e., thermal-charge density, we describe the fine micro-thermal structures of the black hole in detail. The RN black hole actually has three different phases, the {\em free} phase, {\em balanced} phase, and {\em interactive} phase. The thermal-charge density plays a role similar to the order parameter, and the back hole undergoes a new phase transition between the {\em free} phase and {\em interactive} phase. We also observe that the competition between the {\em free} phase and the {\em interactive} phase exists for the RN black hole. In the {\em balanced} phase, the temperature reaches its maximum value. For the extreme RN black hole, the whole system is completely in the {\em interactive} phase and the temperature equals to zero. Furthermore, when the Bogomol'nyi boundary is saturated, i.e., $M=q$, we get an interesting constant, see Eq.~(\ref{rt}). Meanwhile, we provide the thermodynamic micro-mechanism for the formation of the naked singularity for the RN black hole. It is because of the strong attraction interaction when $M<q$ that leads to the whole system becomes a naked singularity.

When $\sigma=0$, i.e. $Q=0$, the situation degenerates to that of the Schwarzschild black hole. For this black hole, its first law of thermodynamics is $d U_{_{\text{Schwarzschild}}}=T_{_{\text{Schwarzschild}}} d S_{_{\text{Schwarzschild}}}$ and the coordinate space of the thermodynamic geometry has only one quantity $S_{_{\text{Schwarzschild}}}$. This renders the metric of the thermodynamic geometry singular, and consequently micro information of the associated black hole is not revealed from the thermodynamic geometry. From this point of view, it is quite possible that the coordinate space of the thermodynamic geometry of the Schwarzschild black hole is incomplete, and we have to analyze some of its micro-behavior with the help of the results of other black holes, like the RN black hole we are currently concerned about. According to Eqs.~(\ref{temperature}) and~(\ref{curvature}), at $Q=0$, we can obtain the expression of the thermodynamic scalar curvature with respect to the temperature for the Schwarzschild black hole
\begin{eqnarray}\label{Schwarzschild}
R_{_{\text{Schwarzschild}}}=16\pi T_{_{\text{Schwarzschild}}}^2.
\end{eqnarray}

From above formula, we see that the thermodynamic scalar curvature is positive, i.e. $R_{_{\text{Schwarzschild}}}>0$ which may be related to the information of repulsive interaction between black hole molecules for the Schwarzschild black hole. Since the black hole is not charged, there will be no electromagnetic interaction, so we can speculate that the repulsion interaction of the black hole is likely to be short-range repulsion between molecules. Some details need to be further analyzed and discussed.

Finally, we need to emphasize several points about the calculation contents of the current paper again. First, there is a general lore that if one knows the
microscopic dynamics of a system, its thermodynamical properties could be derived from statistic physics of the system, while the inverse process does not hold in general, namely one cannot know the micro-dynamics of the system from its thermodynamics. However, for black hole systems, because there does not exist a complete theory of quantum gravity, although the most likely candidate theories---string theory and loop quantum gravity theory---have achieved good results to some extent, the exploration on the microscopic structure of black holes is bound to some speculative assumptions. Meanwhile it is still unclear about the constituents of black holes. Hence, comparing to the methods of studying the usual statistical models, we think that the inverse process mentioned above may be feasible for black hole system as a primary description. Second, due to the well-established black hole thermodynamics, as an analogy analysis and a primary description of the micro-behavior of black holes, one can think that the empirical observation, i.e., a positive (or negative) thermodynamic scalar curvature implies a repulsive (or attractive) interaction and a vanishing thermodynamic scalar curvature corresponds to no interaction, can provide some information about interactions of black holes phenomenologically or qualitatively. Third, the Ruppeiner thermodynamic geometry is based on the Hessian matrix about the black hole entropy, in which the thermodynamic potential plays an important role in the second-order partial differential of other independent thermodynamic quantities. This leads to a completely different result when $q$ or $q^2$ is used as an independent thermodynamic quantity. Maybe when we treat the charge $q$ as a thermodynamic quantity, some information may be cancelled each other, while when we treat the charge square $q^2$ as a thermodynamic quantity, some information can be displayed, see the Appendix for the comparison of thermodynamic metric of the RN black hole under the two schemes.

Furthermore, we hope that our current analysis can be extended to other types of black holes, especially those with AdS background, where we can predict that the AdS background will put the black hole in a new phase. As has been reported in literatures\cite{Wei2015,Wei2019a,Wei2019b,Wei2019c,Miao2018,Miao2019}, the Schwarzschild AdS black hole has attractive interaction, while the charged AdS black hole has both repulsion and attraction. These issues also need to be further explored in the future.

\section*{Acknowledgments}
The financial supports from National Natural Science Foundation of China (Grant Nos. 11947208, 11947301, and 11605137), Major Basic Research Program of Natural Science of Shaanxi Province (Grant No. 2017KCT-12), Scientific Research Program Funded by Shaanxi Provincial Education Department (Program No.18JK0771) are gratefully acknowledged. This research is supported by The Double First-class University Construction Project of Northwest University. The authors would like to thank the anonymous reviewers for the helpful comments that indeed greatly improve this work.

\appendix
\section{Comments on thermodynamic curvature of RN black hole}\label{app}
In Refs.\cite{Aman2003,Aman2006a,Aman2006b} the thermodynamic metric of the RN black hole in the coordinate space $\{S,q\}$ is
\begin{equation}
  \begin{split}
d l^2_q=\frac{2S(S-\pi q^2)}{3\pi q^2-S}d S^2 &+\frac{4\pi q}{\pi q^2-S}dS dq\\
&+\frac{4\pi S}{S-\pi q^2}d q^2.
\end{split}
\end{equation}
With the new coordinate
\begin{equation}
u=\frac{\sqrt{\pi}q}{S}, \qquad -1\leq u \leq 1,
\end{equation}
the above thermodynamic metric reads as in diagonal form
\begin{equation}\label{qmetric}
d l^2_q=-\frac{1}{2S}d S^2+\frac{4S}{1-u^2}d u^2.
\end{equation}
Then under the proper coordinate transformations
\begin{equation}
\tau=\sqrt{2S}, \quad \sin\frac{\sigma}{\sqrt{2}}=u,
\end{equation}
and
\begin{equation}
t=\tau\cosh\sigma, \quad x=\tau\sinh\sigma,
\end{equation}
the thermodynamic metric is finally turned into a Minkowski one
\begin{equation}
d l^2_q=-d \tau^2+\tau^2 d \sigma^2=-d t^2+d x^2,
\end{equation}
which impiles a vanishing scalar curvature in terms of the coordinate space $\{S,q\}$.

In our work, the thermodynamic metric of the RN black hole in the coordinate space $\{S,Q\}$ is
\begin{equation}
d l^2=\frac{2S(S-\pi Q)}{3\pi Q-S}d S^2+\frac{\pi}{\pi Q-S}dS dQ.
\end{equation}
Using the above introduced coordinate $u$, we can write the thermodynamic metric as
\begin{equation}
d l^2=-\frac{1}{2S}d S^2-\frac{2u}{1-u^2}d S d u.
\end{equation}
Compared with Eq.~(\ref{qmetric}), we can clearly see that because of the introduction of $Q=q^2$ and its conjugated quantity $\Psi=1/(2r_h)$ (it is only a function of the horizon radius $r_h$), the thermodynamic metric we obtain is always off-diagonal. Hence the Ruppeiner geometry in our present scheme is non-trivial. In this way, we can get a non-zero thermodynamic curvature, and then realize that the RN black hole is an interaction system.

\end{document}